\documentclass[dvipdfmx]{pasj01}
\Received{$\langle$reception date$\rangle$}
\Accepted{$\langle$acception date$\rangle$}
\Published{$\langle$publication date$\rangle$}
\usepackage{siunitx}
\usepackage{wasysym}
\usepackage{threeparttable}
\usepackage{comment}
\usepackage{array}
\usepackage{url}
\usepackage{mathptmx}

\usepackage[switch,mathlines]{lineno}

\begin{document}

\title{ Plasma diagnostics of supernova remnant 3C\,400.2 by Suzaku observations}
\author{Masataka \textsc{Onuma}\altaffilmark{1,*}, Kumiko K. \textsc{Nobukawa}\altaffilmark{1,*}, Masayoshi \textsc{Nobukawa}\altaffilmark{2}, Shigeo \textsc{Yamauchi}\altaffilmark{3}, Hideki \textsc{Uchiyama}\altaffilmark{4},}
\altaffiltext{1}{Faculty of Science and Engineering, Kindai University, 3-4-1 Kowakae, Higashi-Osaka, 577-8502, Japan}
\altaffiltext{2}{Department of Teacher Training and School Education, Nara University of Education, Takabatake-cho, Nara, 630-8528, Japan}
\altaffiltext{3}{Faculty of Science, Nara Women's University, Kitauoyanishi-machi, Nara, Nara 630-8506, Japan}
\altaffiltext{4}{Faculty of Education, Shizuoka University, 836 Ohya, Suruga-ku, Shizuoka, Shizuoka, 422-8529, Japan}
\email{kumiko@phys.kindai.ac.jp}


\maketitle

\begin{abstract}
We report a result of plasma diagnostics of the supernova remnant (SNR) 3C\,400.2, which has been reported to have a recombining plasma (RP) by previous studies. 
For careful background estimation, we simultaneously fitted spectra extracted from the SNR and background regions and evaluated the SNR emission contaminating the background-region spectrum as well as the background emission in the source-region spectrum. The SNR emission is explained by the collisional ionization equilibrium plasma originating from the interstellar medium and the ionizing plasma originating from the ejecta, in contrast to the previous studies. 
In addition, we found an unidentified X-ray source near the SNR, Suzaku\,J1937.4$+$1718, which is accompanied by an emission line at $\sim4.4$~keV with the 2.8$\sigma$ confidence level. Since there is no striking atomic line at the energy in the rest frame, Suzaku\,J1937.4$+$1718 can be an extragalactic object with a redshifted Fe line.   
\end{abstract}


\section{Introduction}
3C\,400.2 is a Galactic supernova remnant (SNR) located at a distance of $2.8\pm0.8$~kpc (\cite{Giacani98}) and its age is estimated to be $10^4$--$10^5$ years old (\cite{Long91}; \cite{Rosado83}). Its radio morphology shows two overlapping circular shells with diameters of \ang{;14;} and \ang{;22;} (\cite{Dubner94}).
Its optical appearance is shell-like with a radius of approximately $\ang{;8;}$ (\cite{Winkler93}; \cite{Ambrocio06}). 
\citet{Giacani98} reported that dense H~{\emissiontype I} gas overlaps 
 with the radio shell in the northwest of the SNR. 
\citet{Ergin17} found an expanding shell in the position-velocity diagram of H~{\emissiontype I}, the diameter of which is similar to that of the SNR. This indicates the interaction between the H~{\emissiontype I} cloud and the SNR shell. 
Although there is no dense atomic and molecular gas inside the shell, GeV gamma-ray emission was found from the inside (\cite{Ergin17}), and the mechanism of the gamma-ray emission is debatable. 
The Einstein Image Proportional Counter (IPC) and the ROSAT Position Sensitive Proportional Counter (PSPC) revealed an X-ray peak at the center of 3C\,400.2, which was correlated with neither the radio features nor the optical filaments (\cite{Long91}; \cite{Saken95}), and thus 3C\,400.2 was classified into mixed-morphology SNRs \citep{Rho98}. The X-ray spectrum obtained by ASCA was well explained by a thin thermal plasma in collisional ionization equilibrium (CIE) with the temperature of $k T_{\rm e}$=0.5--0.8~keV \citep{Yoshita01}. 

A recombining plasma (RP) of 3C\,400.2 was reported by most recent X-ray studies conducted by Chandra and Suzaku.  \citet{Broersen15} analyzed the Chandra data and claimed that the X-ray spectrum could be explained by two components: an RP originated from the interstellar medium (ISM) ($k T_{\rm e}\sim0.1$~keV) and an ionizing plasma (IP) originated from the ejecta ($kT_{\rm e}\sim3$~keV). \citet{Ergin17}, on the other hand, obtained a result inconsistent with the Chandra result using the Suzaku data; the X-ray spectrum of the eastern region was explained by a CIE plasma originated from the ISM ($kT_{\rm e}=$0.3--0.5~keV) and an RP originated from the ejecta ($k T_{\rm e}=$0.6--0.7~keV) whereas that of the western region was explained by a CIE plasma of the ISM  ($kT_{\rm e}=$0.3--0.4~keV) plus an IP of the ejecta  ($kT_{\rm e}=$0.7~keV).
This discrepancy between the Chandra and Suzaku results may be due to the differences in the way to estimate the background. \citet{Broersen15} used the standard ACIS-I background files to produce the background spectra, whereas \citet{Ergin17} estimated the background spectra by extracting the data from nearby regions of the same observations as the SNR.  

In this paper, we reanalyzed the Suzaku data of 3C\,400.2 with a careful estimation of the background emission to diagnose the plasma state. We performed a simultaneous fitting of spectra extracted from a source region and nearby background region to take into account the SNR emission contaminating the background-region spectrum. We also report a discovery of an unidentified X-ray source near 3C\,400.2 and its spectral analysis result. The errors in the figures are shown in the 1$\sigma$ level, and the uncertainties are quoted at the 90\% confidence range unless otherwise stated.

\section{Observations and Data Reduction}
Suzaku \citep{Mitsuda07} observed four regions of 3C\,400.2 with the X-ray imaging spectrometer (XIS; \cite{Koyama07}), covering the whole SNR. The XIS is placed on the focal plane of the X-Ray Telescope (XRT; \cite{Serlemitsos07}).  The field of view (FOV) of the XIS is \ang{;17;}.8$\times$\ang{;17;}.8. The XIS consists of four X-ray charge-coupled devices (CCDs). XIS1 is a backside illuminated (BI) device, while XIS0, 2, and 3 are frontside illuminated (FI) devices. XIS2 has been dysfunctional since November 9, 2006. A small fraction of XIS0 has been non-functional since June 23, 2009, due to a micro-meteorite impact. We therefore used the data obtained by the remaining parts of XIS0, XIS1, and XIS3. We downloaded the Suzaku archival data of 3C\,400.2 as well as EMS\,1308 for the background estimation via the Data Archives and Transmission System (DARTS; \url{https://www.darts.isas.jaxa.jp/index.html.en}). The observation log is shown in table \ref{tab:log}.

For extracting images and spectra, we used HEASoft version 6.31 and the calibration database version 2018-06-11. The spectral analysis was carried out with XSPEC version 12.10.0. The effective area of the XRT (ancillary response files; ARF) and the response of the XIS (redistribution matrix files; RMF) were calculated using \texttt{xissimarfgen} (\cite{Ishisaki07}) and \texttt{xisrmfgen}, respectively. The non X-ray background (NXB) was estimated using \texttt{xisnxbgen} (\cite{Tawa08}).

\begin{table*}[h]
\caption{Observation log.}\label{tab:log}
\centering
 \begin{tabular}{ccccccc}
\hline
region name & Obs ID & \multicolumn{2}{c}{pointing direction} & \multicolumn{2}{c}{observation time} & exposure time\\
 & & $l\,(^{\circ})$ & $b\,(^{\circ})$ & Start (UT) & End(UT) & (ks) \\
\hline \hline
3C400.2\, NW & 509068010 & $53^{\circ}.68$ &  $-2^{\circ}$.00 & 2014/4/23 11:08:52 & 2014/4/23 22:35:09 & 21.5 \\
3C400.2\, SW & 509069010 & $53^{\circ}.49$ & $-2^{\circ}.18$ & 2014/4/14 12:40:57 & 2014/4/15 01:45:12 & 24.2 \\
3C400.2\, SE & 509070010 & $53^{\circ}.58$ & $-2^{\circ}.41$ & 2014/4/23 22:36:02 & 2014/4/24 11:25:13 & 25.0 \\
3C400.2\, NE & 509071010 & $53^{\circ}.76$ & $-2^{\circ}.41$ & 2014/4/23 00:41:40 & 2014/4/23 11:08:11 & 20.2 \\
EMS\,1308 &  405028010 & $54^{\circ}.37$ & $0^{\circ}.06$ & 2010/4/27 10:28:40 & 2010/4/27 23:26:20 & 23.9 \\ 
\hline
\end{tabular}   
\end{table*}

\section{Analysis and Results}
\subsection{X-ray Images}
\begin{figure*}
\centering
\includegraphics[scale=0.95]{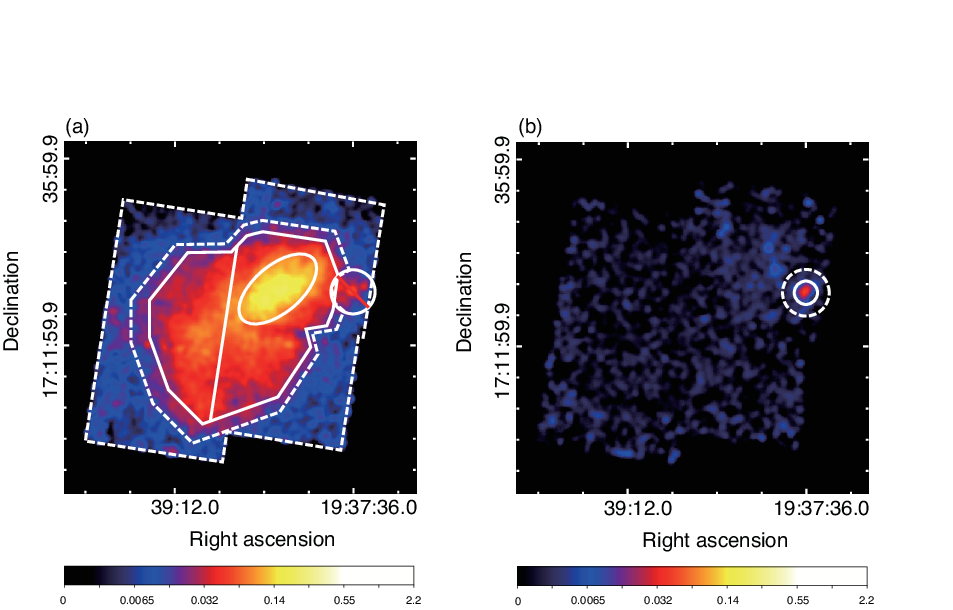}
\begin{flushleft}
   \caption{(a) Suzaku XIS image of 3C\,400.2 in the 0.5--3.0~keV energy band. The NXB is subtracted, and the vignetting effect is corrected. The images of XIS0, 1, and 3 are all merged. The whole SNR region is enclosed by a solid polygon, and it is divided into two regions by a solid line, the ``east'' and ``west'' regions. An ellipse indicates the ``bright'' region. The ``dim'' region is the whole SNR region excluding the bright region. The BGD region is surrounded by white dashed lines. A point-like source, which is indicated by a circle, is excluded from the SNR and BGD spectra.  (b) Same as (a), but the 3.0--5.0~keV energy-band image. The point-like source spectrum and the background spectrum are extracted from the solid circle and an annular region enclosed by the solid and dashed circles, respectively. \label{fig:image} } 
\end{flushleft}

\end{figure*}

Figure \ref{fig:image} shows X-ray images of 3C\,400.2 in the 0.5--3.0 keV and 3.0--5.0 keV energy bands. The NXB was subtracted, and the vignetting effect was corrected. The images of XIS0, 1, and 3 were all merged. We see a diffuse emission due to the SNR in the low energy band, while little X-ray emission other than a point-like source in the high energy band. The coordinate of the point-like source was $(\alpha, \delta)_{\rm J2000}= (\timeform{19h37m36s.13}, \timeform{17D18'53''.82})$ with a positional uncertainty of $19''$ \citep{Uchiyama08}.  It has not been reported in any catalog so far. Then we named this object Suzaku J1937.4$+$1718.

\subsection{Background estimation}
We extracted spectra from the whole SNR region and the background region (hereafter, the BGD region), respectively. The former region is enclosed by a solid polygon, and the latter region is surrounded by the dashed lines in figure \ref{fig:image}(a). 
The FI (XIS0 and 3) spectra were merged, and the FI and BI spectra were simultaneously fitted. 

The X-ray emission of the SNR covers most of the FOV of the four 3C400.2 pointings and it can contaminate the BGD-region spectrum due to the outer skirt of the point spread function (PSF) of the XRT. 
Subtracting the contaminated BGD spectrum from the SNR spectrum results in underestimating the SNR emission flux, and also can affect the best-fit parameters.  
Hence, we conducted simultaneous fitting of the whole SNR and BGD spectra and evaluated the SNR and the background emission fluxes in each region. Simultaneous fitting of source and background spectra has been conducted by several previous studies (\cite{Konami12}, \cite{Ono19}, \cite{Yamauchi23}).  

We generated two ARFs for the whole SNR spectrum: one is for the SNR emission (as an extended source) observed in the whole region, and the other is for the background emission (as a uniform sky) observed in the whole region. Applying the two different ARFs to the two models follows the XSPEC manual described in \url{https://heasarc.gsfc.nasa.gov/xanadu/xspec/manual/node40.html}.
Similarly, we made other two ARFs for the BGD spectrum: one is for the SNR emission contaminating the BGD region due to the outskirt of the PSF, and the other is for the background emission observed in the BGD region. 

Since 3C\,400.2 is located in the inner region of our Galaxy, the background emission is composed of the foreground emission (FE; e.g. the local hot bubble), the Galactic ridge X-ray emission (GRXE), and the cosmic X-ray background (CXB). 
Referring to \citet{Uchiyama13}, we adopted the following model for the background emission; Abs1 $\times$ FE $+$ Abs2 $\times$(HP$_{\rm GRXE}$ $+$ LP$_{\rm GRXE}$ $+$ CM $+$ Abs2 $\times$ CXB). 
``Abs1'' and ``Abs2'' mean the interstellar absorptions for the FE and the GRXE, respectively, and the hydrogen column densities $N_{\rm H}$ of the individual components are free parameters. 
``FE'' consists of two-temperature CIE plasmas, and we fixed the parameters other than the normalizations to those of \citet{Uchiyama13}. 
The GRXE components ``HP'' and ``LP'' are high- and low-temperature plasmas, respectively. We fixed their parameters other than the normalizations to those of \citet{Uchiyama13}.  ``CM'' comes from the cold matter and is composed of the neutral Fe K$\alpha$ and K$\beta$ lines and the power-law emission associated with them. 
We fixed all the parameters of ``CM''  (the centroids, the equivalent widths, and the flux ratio of the K$\alpha$ and K$\beta$ lines) other than the flux of the Fe K$\alpha$ line to those of \citet{Uchiyama13}. ``CXB'' is expressed as a power-law function, whose photon index and flux were fixed to those of \citet{Kushino02}.

We found that the simultaneous fitting resulted in large uncertainties of the flux ratio of ``FE'', ``HP'', and ``LP'' because the fluxes of the components were degenerated.  
Then we referred to a spectrum of the nearby region (EMS\,1308) to constrain the flux ratio of the background components. 
The nearby-region spectrum was well explained by the background model. 
The Galactic longitude of the FOV of the nearby region is almost the same as the SNR coordinate, but the Galactic latitude is lower than the SNR's (see table \ref{tab:log}).
Since the HP and LP fluxes depend on the Galactic coordinates, 
we scaled the best-fit fluxes of ``HP'' and ``LP'' in the nearby region to that of the central position of 3C\,400.2 by the scale heights obtained by \citet{Yamauchi16}. 

We can now conduct the simultaneous fitting of the whole SNR and BGD spectra.  
The X-ray spectra in the high-energy band were dominated by the CXB. This is consistent with the X-ray image in the high-energy band (figure \ref{fig:image}b), where no structures other than the point-like source are found. In addition, the ratio of the NXB to the signal increases in the high-energy band. Therefore, the energy band above 5~keV is discarded in the following analysis. 
We will describe the detail and fitting result in the next section.

\begin{table}
\begin{threeparttable}[h]
 \caption{Best-fit parameters of BGD region spectrum.\label{tab:bgd}}
           \centering
           \begin{tabular}{llc}
        \hline
       Component & Parameter & Value \\
       \hline
       Abs1 & $\it{N}_{\rm H}$ ($\SI{E22}{\per\square\cm}$) & $1.03^{+0.01}_{-0.04}$ \\
       \hline
       FE (CIE) & $k T_{\rm e}$ (keV) & 0.09 (fixed)\tnote{$\parallel$} \\
        & Abundance\tnote{*} & 0.05 (fixed)\tnote{$\parallel$} \\  
        & Norm\tnote{\dag} & $28\pm4$ \\
       \hline
       FE (CIE) & $k T_{\rm e}$ (keV) & 0.59 (fixed)\tnote{$\parallel$} \\
        & $\rm{Z_{Ne}}$\tnote{*} & $0.25^{+0.07}_{-0.10}$ \\
        & $\rm{Z_{Mg}}$\tnote{*} & $0.15^{+0.04}_{-0.03}$ \\
        & $\rm{Z_{Other}}$\tnote{*} & 0.05 (fixed)\tnote{$\parallel$} \\
        & Norm\tnote{\dag}& ($5.0\pm0.5$)$\times10^{-3}$ \\
       \hline
       Abs2 & $\it{N}_{\rm{H}}$ ($\SI{E22}{\per\square\cm}$) & $2.70^{+0.21}_{-0.18}$ \\
       \hline
       HP (CIE) & $k T_{\rm e}$ (keV) & 6.64 (fixed)\tnote{$\parallel$} \\
        & $\rm{{Z}_{Ar}}$\tnote{*}& 1.07 (fixed)\tnote{$\parallel$} \\
        & $\rm{Z_{Other}}$\tnote{*} & 0.81 (fixed)\tnote{$\parallel$} \\
        & Norm\tnote{\dag} & 1.3$\times10^{-4}$\,(fixed) \\
       \hline
       LP (CIE) & $k T_{\rm e}$ (keV) & 1.33 (fixed)\tnote{$\parallel$} \\
        & $\rm{{Z}_{Ar}}$\tnote{*}& 1.07 (fixed)\tnote{$\parallel$} \\
        & $\rm{Z_{Other}}$\tnote{*}& 0.81 (fixed)\tnote{$\parallel$} \\
        & Norm\tnote{\dag} & 2.0$\times10^{-3}$ (fixed) \\
       \hline
       CM (Fe~{\emissiontype I} K$\alpha$) & centroid (keV) & 6.40 (fixed)\tnote{$\parallel$} \\
        & Norm\tnote{\ddag}& $<5.2\times10^{-8}$\\
       CM (Fe~{\emissiontype I} K$\beta$) & centroid (keV) & 7.06 (fixed)\tnote{$\parallel$} \\
        & Norm\tnote{\ddag} & $\rm{Norm}_{6.4}\times0.125$ \\
       CM (Power law) & $\Gamma$ & 2.13 (fixed)\tnote{$\parallel$} \\
        & $EW_{6.4}$ (eV) & 547 (fixed)\tnote{$\parallel$} \\
       \hline
       CXB (Power law) & $\Gamma$ & 1.4 (fixed)\tnote{$\sharp$} \\
        & Norm\tnote{\S} & 8.6$\times10^{-5}$ (fixed)\tnote{$\sharp$} \\
         \hline
         \end{tabular}
      \begin{tablenotes}
      \item[*] Relative to the solar values in \citet{Anders89}.
      \item[\dag] Defined as $\frac{10^{-14}}{4\pi{D_A} ^2}\int n_{\rm e} n_{\rm H} dV$. $D_A$ is the angular diameter distance to the source. $dV$ is the volume element, and $n_{\rm e}$ and $n_{\rm H}$ are the electron and hydrogen densities, respectively. The unit is $\si{cm^{-5}}$.
      \item[\ddag] The unit is $\rm{photons}~\rm{cm}^{-2}~\rm{s}^{-1}$.
      \item[\S] The unit is  $\rm{photons}~\rm{keV}^{-1}~\rm{cm}^{-2} ~\rm{s}^{-1}$ at 1~keV.
      \item[$\parallel$] The values are fixed \citet{Uchiyama13}.
      \item[$\sharp$]The values are fixed \citet{Kushino02}.
      \end{tablenotes}
      \end{threeparttable}
      \end{table}

\subsection{Simultaneous fitting of the whole SNR and BGD spectra}
 In the simultaneous fitting described below, we found that there are residuals in the BGD spectrum at the Ne and Mg K-shell lines of $\sim1.0$~keV and $\sim 1.3$~keV, respectively,  regardless of the SNR emission models we tried. These lines are dominated by the ``FE'' component with the temperature of 0.59~keV. \citet{Uchiyama13} examined abundances of the GRXE spectrum above 2~keV, and thus Ne and Mg abundances were not constrained. We let the Ne and Mg abundances of the ``FE'' component free. 
 
 We first assumed that the SNR emission comes from a one-temperature IP, and adopted the \texttt{phabs}$\times$\texttt{vnei} model in XSPEC. All the parameters of the SNR model other than abundances were set as free parameters. The abundances of Ne, Mg, Si, S, Ar, Ca, Fe, and Ni were free, while the other abundances were fixed to the solar ones. Since the line intensities of Ar, Ca, Fe, and Ni are not strong enough, the abundances of Ar and Ca, and those of Fe and Ni were linked, respectively. The best-fit model could not explain the SNR spectra in the low-energy band, especially around $\sim$1.3~keV and $\sim$1.9~keV. The reduced $\chi^2$ was 1.85 (d.o.f.=613), which is not acceptable.  
Then we adopted a two-temperature IP model;  \texttt{phabs}$\times$(\texttt{vnei}$+$\texttt{vnei}). Assuming that the low- and high-temperature components are of the ISM and ejecta origins, respectively, we allowed the Ne, Mg, Si, S, Ar, Ca, Fe, and Ni abundances of the high-temperature component to be free, while the other abundances, including those of the low-temperature component, were fixed to the solar ones.  
The reduced $\chi^2$ of the fitting was 1.69 (d.o.f.=610), which is still unacceptable. We found residuals at $\sim0.8$~keV and $\sim1.2$~keV. It would be due to the incomplete atomic data for Fe L-shell emission lines in the XSPEC model (e.g., \cite{Nakasima13}). We added two Gaussians to represent these features, which decreased the reduced $\chi^2$ to 1.15 (d.o.f.=606). 
The best-fit ionization timescale ($\tau=n_{\rm e}t$) of the low-temperature component was $7.3\times\SI{e12}{cm^{-3}.s}$. This value indicates that the plasma has already reached to the CIE state. We finally adopted the SNR model of \texttt{phabs}$\times$(\texttt{vapec}$+$\texttt{vnei}$+$\texttt{gaussian}$+$\texttt{gaussian}), where we regard the low-temperature component as the CIE plasma. 
Figures \ref{fig:snr_spec}(a) and (b) show the simultaneous fitting results of the BGD and the whole SNR spectra, respectively. 
The simultaneous fitting revealed that the background emission contributes to 10\% of all the X-ray emissions in the whole SNR region, while the SNR emission contaminating in the BGD region accounts for 5.8\% of the X-ray emission there.
Tables \ref{tab:bgd} and \ref{tab:parameter} show the best-fit parameters of the BGD and the whole SNR spectra, respectively.

\citet{Broersen15} analyzed the Chandra data and explained a 3C\,400.2 spectrum with a two-temperature plasma model: an RP originating from the ISM ($k T_{\rm e}\sim0.1$~keV) plus an IP originating from the ejecta ($kT_{\rm e}\sim3$~keV).
To examine whether the model is valid, we fitted the whole SNR spectra with the model referred by \citet{Broersen15}, adding the Fe-L lines, \texttt{phabs}$\times$(\texttt{vrnei}$+$\texttt{vrnei}$+$\texttt{gaussian}$+$\texttt{gaussian}) in XSPEC. The fitting resulted in positive residuals at $\sim$0.7~keV and the He-like Si line, and the model was rejected by the reduced-$\chi^2$ value of 3.92 (d.o.f.=404).

\subsection{Spectral analysis of the bright and dim regions}
The X-ray image of 3C\,400.2 (figure \ref{fig:image}a) shows a bright emission in the northwestern region surrounded by a relatively dim emission. To investigate the difference in the plasma parameters depending on the brightness, we divided the whole SNR region into two regions, ``bright'' (the eclipse with the radii of $3.0'$ and $6.0'$ in figure \ref{fig:image}a) and ``dim'' (the whole SNR region excluding the eclipse), and extracted spectra from each region. 

Similar to the simultaneous fitting of the whole SNR and the BGD spectra, we fitted a source spectrum with the two ARFs; one for the SNR emission model and the other for the background model. We used the background model that was obtained by the simultaneous fitting of the whole SNR and BGD spectra, with all the parameters of the model fixed to the best-fit ones in the simultaneous fitting, except for the normalizations, which were scaled by the sky areas of the source region. 

The SNR emission model is the same as the best-fit one in the simultaneous fitting, \texttt{phabs}$\times$(\texttt{vapec}$+$\texttt{vnei}$+$\texttt{gaussian}$+$\texttt{gaussian}) in XSPEC.  We allowed the hydrogen column density $\it{N}_{\rm H}$, the temperatures $k T_{\rm e}$, the ionization time scale $\tau$, the abundances of Ne, Mg, Si, S, Ar, Ca, Fe, and Ni (in $\texttt{vnei}$), and the normalizations to be free. 
The model explained the spectra of both the regions well. 
The obtained reduced--$\chi^2$ values of the ``bright'' and ``dim'' spectra were 0.94 (d.o.f.=409) and 1.06 (d.o.f.=394), respectively. The spectra are shown in figures \ref{fig:snr_spec}(c) and (d), and the best-fit parameters are listed in table \ref{tab:parameter}.

\subsection{Spectral analysis of the east and west regions}

The Suzaku data we analyzed was previously studied by \citet{Ergin17}, who divided the SNR region into four regions, NW, SW, SE, and NE, and found an RP in the SE and NE regions. To examine the RP scenario, we extracted spectra from the ``east'' region (roughly a sum of SE and NE) as well as from the ``west'' region (roughly a sum of NW and SW). The source regions are shown in figure \ref{fig:image}(a).
We used the background model that was obtained by the simultaneous fitting of the whole SNR and BGD spectra, and all the parameters were fixed to the best-fit ones in the previous simultaneous fitting, except for the normalizations, which were scaled by the sky areas. We fitted each region's spectra with the two ARFs; one is for the SNR emission model and the other for the background model.

\begin{figure*}[h]
\centering
\includegraphics[width=13cm]{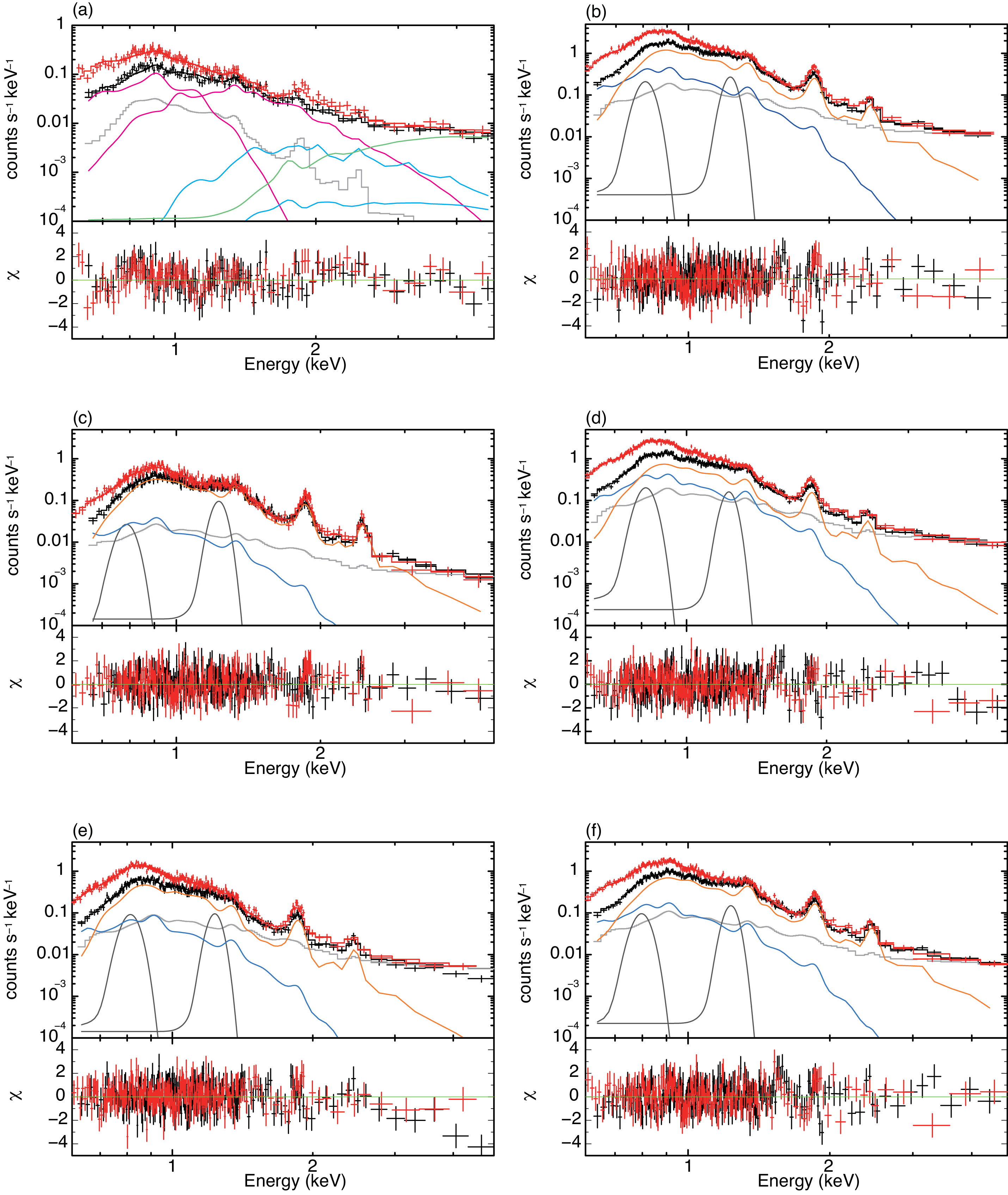}
\begin{flushleft}
 \caption{(a) BGD spectra with the best-fit background model. The magenta, cyan, green, and gray lines indicate ``FE'', ``LP'' and ``HP'', ``CXB'', and the contamination of the SNR 3C\,400.2, respectively. (b) Whole SNR spectra with the best-fit model. The blue, orange, black, and gray lines show the low-temperature plasma, the high-temperature plasma, the Fe-L lines, and the background model, respectively. The BGD spectra (a) and whole SNR spectra (b) are simultaneously fitted. (c) ``dim'' region spectra. The color indication is the same as (b).  (d) Same as (c), but for the ``bright'' region spectra. (e) Same as (c), but for the ``east'' region. (f) Same as (c), but for the ``west'' region. The black and red data show the FI and BI spectra, respectively. \label{fig:snr_spec}} 
\end{flushleft}

\end{figure*}

We adopted the same SNR emission model as was obtained by the fitting of the whole region spectrum, \texttt{phabs}$\times$(\texttt{vapec}$+$\texttt{vnei}$+$\texttt{gaussian}$+$\texttt{gaussian}) in XSPEC. The free parameters are the same as in the fitting of the ``bright'' and ``dim'' spectra. 
Our model explained the ``east'' and ``west'' spectra well. The reduced--$\chi^2$ values of the ``east'' and ``west'' spectra were 1.07 (d.o.f.=406) and 1.12 (d.o.f.=391), respectively. The spectra are shown in figures \ref{fig:snr_spec}(e) and (f), and the best-fit parameters are listed in table \ref{tab:parameter}.
We also fitted the ``east'' spectrum with the model referred by \citet{Ergin17}, adding the Fe-L lines, \texttt{phabs}$\times$(\texttt{vrnei}$+$\texttt{vrnei}$+$\texttt{gaussian}$+$\texttt{gaussian}) in XSPEC. The model was rejected by the reduced-$\chi^2$ value of 3.60 (d.o.f.=419); there were remarkable positive residuals in the Si and S lines.

\begin{table*}[ht]
\begin{threeparttable}[h]
\caption{Best-fit parameters of 3C\,400.2.\label{tab:parameter}}
            \centering
           \begin{tabular}{p{14mm}p{18mm}m{23mm}m{23mm}m{23mm}m{23mm}m{23mm}}
        \hline
       Component & Parameter & \multicolumn{5}{c}{Value} \\
        \hline
        & & whole & bright & dim & east & west \\
       \hline
       Absorption & $\it{N}_{\rm H}$\tnote{*} & $0.54^{+0.01}_{-0.02}$ & $0.54^{+0.02}_{-0.03}$ & $0.57^{+0.03}_{-0.02}$ & $0.50^{+0.02}_{-0.01}$ & $0.52^{+0.01}_{-0.02}$ \\
       \hline
       CIE & $k T_{\rm e}$ (keV)& $0.23^{+0.01}_{-0.02}$ & $0.20\pm0.03$ & $0.24\pm0.01$ & $0.20^{+0.03}_{-0.03}$ & $0.22^{+0.01}_{-0.03}$ \\
        & Abundance \tnote{\dag}& 1.0 (fixed) & 1.0 (fixed) & 1.0 (fixed) & 1.0 (fixed) & 1.0 (fixed) \\  
        & Norm\tnote{$\parallel$} & $14\pm2$ & $2.8^{+4.1}_{-1.7}$ & $13\pm2$ & $5.5^{+2.4}_{-1.3}$ & $7.3^{+1.5}_{-0.6}$ \\
       \hline
       IP & $k T_{\rm e}$ (keV) & $0.75\pm0.01$ & $0.81^{+0.01}_{-0.03}$ & $0.75\pm0.02$ & $0.69^{+0.02}_{-0.01}$ & $0.80^{+0.03}_{-0.02}$ \\
        & $\rm{Z_{Ne}}$\tnote{\dag} & $1.5\pm0.1$ & $1.6^{+0.8}_{-0.6}$ & $2.5\pm0.5$ & $3.7^{+1.0}_{-0.8}$ & $1.2\pm0.2$ \\
        & $\rm{Z_{Mg}}$\tnote{\dag} & $2.3^{+0.7}_{-0.4}$ & $4.3^{+2.4}_{-0.9}$ & $3.2\pm0.2$ & $4.6\pm0.5$ & $2.1^{+0.1}_{-0.3}$ \\
        & $\rm{Z_{Si}}$\tnote{\dag} & $2.7\pm0.3$ & $3.7^{+1.6}_{-0.9}$ & $4.6^{+2.2}_{-0.3}$ & $6.9\pm0.7$ & $1.9^{+0.2}_{-0.1}$ \\
        & $\rm{Z_{S}}$\tnote{\dag} & $3.8^{+1.3}_{-1.1}$ & $6.7^{+2.2}_{-2.1}$ & $5.7^{+1.0}_{-0.9}$ & $9.0^{+2.5}_{-2.4}$ & $3.0\pm0.4$ \\
        & $\rm{Z_{Ar, Ca}}$\tnote{\dag} & $<$5.7 & $<$2.3 & $<$10 & $<$6.0 & $<$4.5 \\
        & $\rm{Z_{Fe, Ni}}$\tnote{\dag} & $3.7^{+0.2}_{-0.8}$ & $4.6^{+2.7}_{-1.1}$ & $7.0^{+0.6}_{-0.9}$ & $12\pm1$ & $2.3^{+0.3}_{-0.2}$ \\
        & $\rm{Z_{other}}$\tnote{\dag}& 1.0 (fixed) & 1.0 (fixed) & 1.0 (fixed) & 1.0 (fixed) & 1.0 (fixed) \\
        & $\tau$\tnote{\ddag}& $1.9^{+0.2}_{-0.3}$ & $1.6^{+0.3}_{-0.4}$ & $1.5\pm0.3$ & $1.4^{+0.4}_{-0.2}$ & $1.4^{+0.3}_{-0.2}$ \\
        & Norm\tnote{$\parallel$} & $1.6\pm0.2$ & $0.48^{+0.13}_{-0.09}$ & $0.69^{+0.04}_{-0.12}$ & $0.3\pm0.1$ & $1.7\pm0.1$ \\
         \hline
        $\chi^2$/d.o.f & &  691.21/608=1.14 & 384.59/409=0.94 & 417.93/394=1.06 & 435.31/406=1.07 & 437.98/391=1.12 \\ 
         \hline
         \end{tabular}
 \begin{tablenotes}
      \item[*] The unit is $10^{22}~\si{\per\square\cm}$ .
      \item[\dag] Relative to the solar values in \citet{Anders89}.
      \item[\ddag] Ionization timescale. The unit is $ 10^{11}~\si{\cm^{-3}.s}$.
      \item[$\parallel$] Defined as $\frac{10^{-14}}{4\pi{D_A} ^2}\int n_{\rm e} n_{\rm H} dV$. $D_A$ is the angular diameter distance to the source. $dV$ is the volume element, and $n_{\rm e}$ and $n_{\rm H}$ are the electron and hydrogen densities, respectively. The unit is $10^{-3}~\si{\cm^{-5}}$.
  \end{tablenotes}
      \end{threeparttable}
      \end{table*}

\subsection{Spectral analysis of Suzaku J1937.4$+$1718}
We extracted FI and BI spectra from the $\timeform{1'.5}$ circle around the unidentified source, Suzaku J1937.4$+$1718,  as well as from the annular region around the circle with the radius of $\timeform{3'.0}$ for the background subtraction (figure \ref{fig:image}b). 

\begin{figure}[h]
\centering
 \includegraphics[scale=0.3]{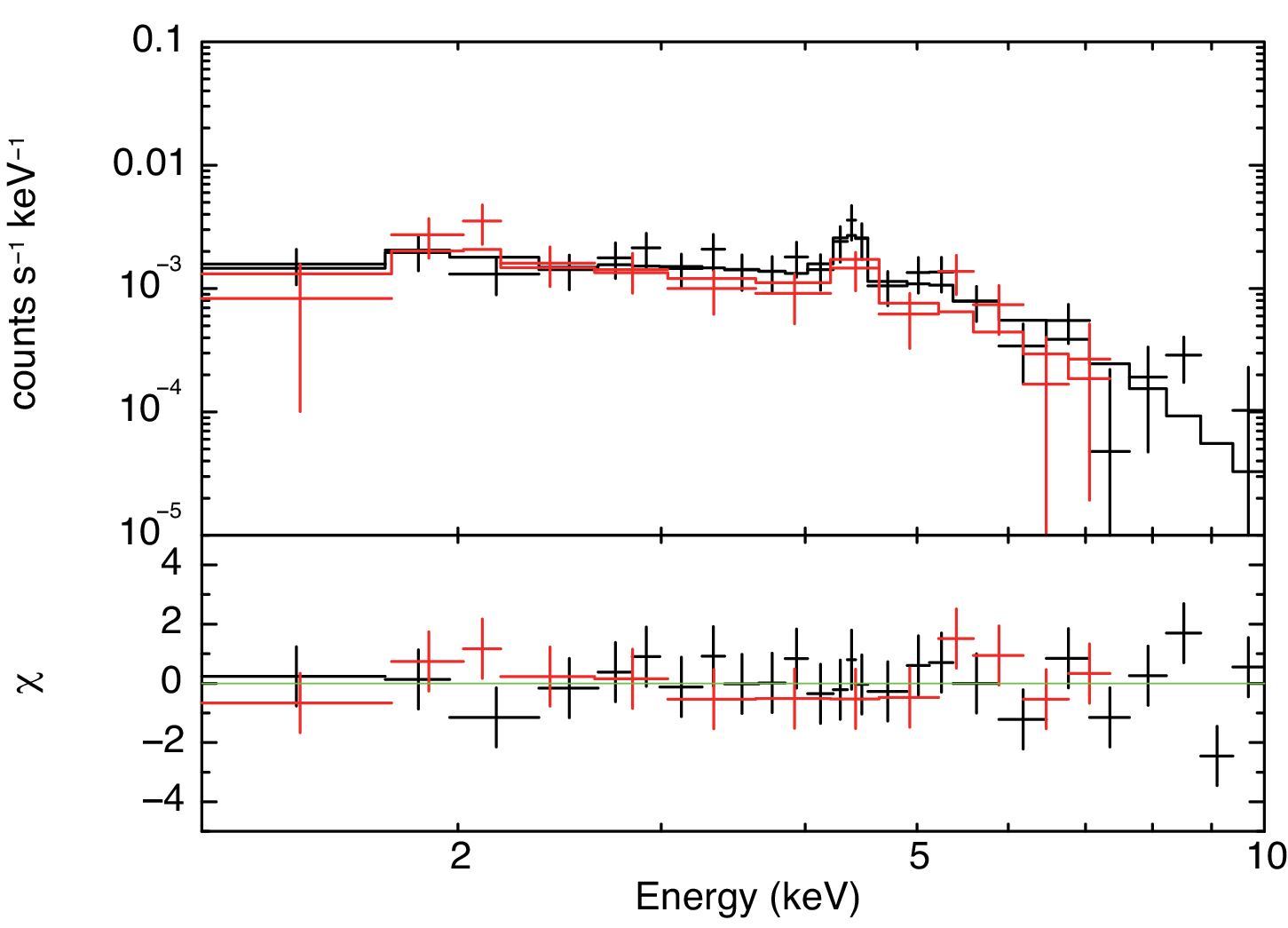}
 \begin{flushleft}
  \caption{Background-subtracted spectra Suzaku J1937.4$-$1718  fitted with the absorbed CIE model (model~B). The black and red data represent the FI and BI spectra, respectively. \label{fig:ps_spec} }
 \end{flushleft}
\end{figure}

We fitted the background-subtracted spectra with an absorbed power-law. 
The spectra show line-like residuals at $\sim$4.4~keV. We then fit the spectra with the model consisting of an absorbed power-law plus a gaussian (model~A). 
The line feature at $4.4\pm0.1$~keV was detected with the significance level of 2.8$\sigma$ and the line equivalent width (EW) is measured to be $340^{+560}_{-270}$~eV (table \ref{tab:srcA}). Since there is no striking atomic line at the rest energy around $\sim$4.4~keV, the line feature could be due to a red-shifted Fe line; a neutral Fe line at 6.4~keV or He-like Fe line at 6.7~keV in the rest frame. If the former is the case, the redshift is calculated to be $z=0.46^{+0.03}_{-0.04}$.  
Assuming the latter case, we also fit the spectra with an absorbed CIE plasma model (model~B). We obtained the electron temperature $k T_{\rm e}$ of $8.2^{+8.0}_{-2.9}$~keV and the redshift of $z=0.55^{+0.05}_{-0.04}$. The best-fit parameters of model~A and model~B are shown in table \ref{tab:srcA}. The spectra are shown in figure \ref{fig:ps_spec}.
The flux of the point-like source in the 0.2--16~keV band is measured to be $6.7^{+0.3}_{-2.1} \times 10^{-13}$~erg~s$^{-1}$~cm$^{-2}$ and $6.0^{+0.8}_{-2.6} \times 10^{-13}$~erg~s$^{-1}$~cm$^{-2}$  
for model~A and model~B, respectively. 

\begin{table}[h]
\begin{threeparttable}[h]
\caption{Best-fit parameters of Suzaku J1937.4$+$1718.\label{tab:srcA}}
\centering
           \begin{tabular}{llc}
           \hline
       Model & Parameter & Value \\
        \hline
       \multicolumn{3}{c}{model A} \\
       Absorption & $\it{N}_{\rm H}$  ($10^{22}$~cm$^{-2}$) & $2.8^{+1.4}_{-1.0}$ \\
       Power law & $\Gamma$ & $2.2^{+0.6}_{-0.5}$ \\
        & Norm\tnote{*} & $3.4^{+5.2}_{-1.9}$ \\
       Gaussian & centroid (keV) & $4.4\pm0.1$ \\
        & width (eV) & 0 (fixed) \\
        & Norm\tnote{\dag} & $4.3\pm2.5$ \\
        & $\rm{EW}_{gauss}$ (eV) \tnote{\ddag} & $340^{+560}_{-270}$ \\
       $\chi^2$/d.o.f.  & & 47.23/29=1.63 \\
       \hline
       \multicolumn{3}{c}{model B} \\
       Absorption & $\it{N}_{\rm H}$ ($10^{22}$~cm$^{-2}$) & $2.1^{+1.0}_{-0.7}$ \\
       CIE & $k T_{\rm e}$ (keV) & $8.2^{+8.0}_{-2.9}$\\
        & Abundance\tnote{$\parallel$} & $0.75^{+1.86}_{-0.47}$\\
        & redshift & $0.55^{+0.05}_{-0.04}$\\
        & Norm\tnote{$\sharp$} & $1.1^{+0.7}_{-0.5}$ \\
       $\chi^2$/d.o.f. & & 42.34/29=1.46 \\
       \hline
         \end{tabular}
 \begin{tablenotes}
      \item[*] The unit is  $10^{-4}~\rm{photons}\ 
 \si{keV^{-1}.cm^{-2}.s^{-1}}$ at 1~keV.
      \item[\dag] The unit is $10^{-6}~\rm{photons}\ \si{cm^{-2}.s^{-1}}$.
      \item[\ddag] Equivalent width. 
      \item[$\parallel$] Relative to the solar values in \citet{Anders89}.
      \item[$\sharp$] Defined as $\frac{10^{-14}}{4\pi[D (1+\rm{z})]^2}\int n_{\rm e} n_{\rm H} dV$. $D$ is the distance to the source. $dV$ is the volume element, and $n_{\rm e}$ and $n_{\rm H}$ are the electron and hydrogen densities, respectively. The unit is $10^{-3}~\si{\cm^{-5}}$.
  \end{tablenotes}
      \end{threeparttable}
      \end{table}

\section{Discussion}

\subsection{SNR plasma of 3C\,400.2}
We adapted the whole SNR spectra with the two-temperature model: the low-temperature ($\sim0.2$~keV) CIE plasma plus the high-temperature ($\sim0.8$~keV) IP.  
On the assumption that the former would be of the ISM heated by a blast wave while the latter would be of the ejecta origin, we fixed the abundances of the low-temperature plasma to 1 solar whereas those of Ne, Mg, Si, S, Ar, Ca, Fe, and Ni of the high-temperature plasma were free parameters. The model explained the whole SNR spectra well. Although the abundances of Ar and Ca were only constrained by the upper limit, those of the other elements were obtained to be higher than 1 solar.  

Assuming that the distance of the SNR is $D=2.8$~kpc (\cite{Giacani98}) and the plasma of the whole SNR region is shaped like a sphere with a 12-arcmin radius, the X-ray emitting volume is estimated to be $V=1.1\times10^{59}f$~cm$^{3}$. Here, $f$ is a filling factor. The best-fit normalization of the ejecta component is defined as $\frac{10^{-14}}{4\pi D^2} \int n_{\rm e} n_{\rm H} dV \sim \frac{10^{-14}}{4\pi D^2} n_{\rm e} n_{\rm H} V$,  where $n_{\rm e}$ and $n_{\rm H}$ are the electron and Hydrogen densities, respectively, and is obtained to be  $1.6\times10^{-3}$~cm$^{-5}$ as shown in table~3. On the assumption of the density ratio $n_{\rm e}/n_{\rm H}=1.2$, 
we estimated the averaged electron density of the ejecta to be $n_{\rm e} \sim 0.04 f^{-1/2}~\si{cm^{-3}}$, and then obtained the ionization timescale of $t =\tau /n_{\rm e} \sim 1.5\times10^5 f^{1/2}$~yr for the whole region. This is compatible with the SNR age estimated by the previous studies ($10^4$--$10^5$~yr; \cite{Long91}; \cite{Rosado83}).
The low density of the ejecta preserves the plasma in the ionizing state despite the old age of 3C\,400.2. We calculated the total ejecta mass by $M_{\rm ejecta} \sim \sum_i \eta_i n_{\rm H} Z_i m_i V$, where $\eta_i$, $Z_i$, $m_i$ are the abundance ratio to Hydrogen in the solar abundance \citep{Anders89}, the ejecta abundance relative to the solar ones, and the mass of each element ($i=$He, C, N, O, Ne, Mg, Si, S, Ar, Ca, Fe, and Ni). The obtained ejecta mass is 
$M_{\rm ejecta}\sim4.5 M_{\astrosun}f^{1/2}$, which indicates the core-collapse (CC) supernova origin.  

We estimated the progenitor of 3C\,400.2 using the abundance ratio of the ejecta components. \citet{Katsuda18} investigated a progenitor mass distribution of core-collapse SNRs in our Galaxy and Large and Small Magellanic Clouds and elucidated that the Fe/Si ratio is sensitive to the progenitor's zero-age main-sequence mass $M_{\rm ZAMS}$. According to their classification, the Fe/Si ratio observed in 3C\,400.2, $1.4^{+0.2}_{-0.3}$, indicates the progenitor with $M_{\rm ZAMS} < 15 M_{\astrosun}$, which is the lowest mass group among core-collapse SNRs. In fact, the Fe/Si ratio of the SNR is notably higher compared to other core-collapse SNRs, suggesting significant production of $^{56}$Ni. One possible scenario for the progenitor of 3C\,400.2 is a hydrogen-poor stripped-envelope supernova, which can produce at least three times as much $^{56}$Ni as other classes of core-collapse supernovae can \citep{Anderson19,Afsariardchi21}.

We analyzed the ``bright'' and ``dim'' spectra to see the difference in the plasma parameters depending on the brightness. 
The temperatures of the ISM components as well as the abundances and the ionization time scales of the ejecta components are consistent between the two regions within the 90\% confidence level. 
On the other hand, one of the differences in the parameters is the electron temperature of the ejecta component, which is higher in the ``bright'' region ($0.81^{+0.01}_{-0.03}$~keV) than in the ``dim'' region ($0.75\pm0.02$~keV). 
Another difference is the ISM density. Assuming that the volumes of the bright region as an ellipsoid with radii of $3.0', 3.0'$, and $6.0'$, we estimated the electron density of the ISM component in the ``bright'' region to be $\sim$0.29$f^{-1/2}~\si{cm^{-3}}$, which is about 2.4 times higher than that of the dim region ($\sim$0.12$f^{-1/2}~\si{cm^{-3}}$). 
The previous study (\cite{Ergin17}) found an indication of the interaction between the SNR shell and the H~{\emissiontype I} cloud located in the northwest of the SNR. The higher density of the ISM component in the ``bright'' region would come from the H~{\emissiontype I} gas. 
In summary, a plausible scenario is that 3C\,400.2 is a remnant of a core-collapse supernova (possibly a hydrogen-poor stripped-envelope supernova), which exploded in the H~{\emissiontype I} gas cavity near the ``bright'' region.

\subsection{Comparison with the previous studies}
Our analysis revealed that the 3C\,400.2 spectra are explained by a combination of the CIE plasma and the IP, in contrast to the previous studies which explained the SNR spectra with the RP model. 
We discuss possible causes of the discrepancy between the previous studies and this study.

\citet{Broersen15} reported that radiative recombination continua (RRCs) are seen in the 0.5--1.7~keV band in the Chandra spectrum. In this study, however, the RRCs were not found in the whole SNR spectra. A difference in the analysis between \citet{Broersen15} and this study is the methods of the background estimation; \citet{Broersen15} produced the background spectra by using the standard ACIS-I background files, whereas we conducted the simultaneous fitting.  \citet{Broersen15} found the SNR emission even above 3~keV in the background-subtracted spectra, but we found that the spectra above 3~keV is dominated by the background emission as shown in figure \ref{fig:snr_spec}(b). We consider that the Chandra study would have underestimated the background emission. In fact, \citet{Broersen15} obtained a much higher temperature of $kT_{\rm e}\sim3$~keV than those obtained by \citet{Ergin17} and this study. The higher the plasma temperature, the more insufficient the continuum emission in the low-energy band. We infer that the low-temperature RP of  the \citet{Broersen15} model is required to complement the insufficient continuum with the RRCs.

\citet{Ergin17} reported that the SNR emission in the NE and SE regions is explained by the RP of the ejecta origin as well as the CIE plasma of the ISM origin. On the other hand, we found that the SNR emission in the east region can be explained by the IP of the ejecta origin and the CIE plasma of the ISM origin, despite analyzing the same Suzaku data as the authors did. \citet{Ergin17} found that the spectra of the NW and SW regions are expressed by the CIE plasma and IP, which is consistent with this study. 
The difference between \citet{Ergin17} and this study in the east region is also attributable to the difference in the background estimation. \citet{Ergin17} estimated the background spectra by extracting the data from nearby regions of the same FOVs as the SNR. The nearby-region spectra contain contamination from the 3C\,400.2 emission with the strong line emissions as demonstrated in this study. Therefore, subtracting the nearby-region spectra from the source ones leads to an overestimation of the background emission, especially in the emission lines.     
The RP model was needed in the \citet{Ergin17} study because it would explain the weaker line intensities than those expected from the electron temperature in the IP or CIE plasma model. Meanwhile, since the NW and SW spectra have larger normalizations than the NE and SE ones, the NW and SW spectra were less affected by the background subtraction, and thus they were explained by the same model as this study.

\subsection{Origin of Suzaku J1937.4$+$1718}
We discovered the unidentified source Suzaku J1937.4$+$1718. The hydrogen column density of the source ($N_{\rm H} = 2.8^{+1.4}_{-1.0}\times10^{22}$~cm$^{-2}$ in model~A) is five times as large as that of 3C\,400.2 ($N_{\rm H} = 0.54^{+0.01}_{-0.02}\times10^{22}$~cm$^{-2}$), which implies the extragalactic origin.  
Moreover, we detected the line structure at $E\sim$4.4~keV with the 2.8$\sigma$ significance level in the X-ray spectra. Since there are no striking atomic lines at this energy in the rest frame, the line would be due to a redshifted Fe line, and the source is considered to be an active galactic nucleus (AGN) or a cluster of galaxies (CG). 

Assuming that the source is an AGN and fitting the spectra with the model consisting of an absorbed powerlaw and a gaussian representing a redshifted Fe~{\emissiontype I} K$\alpha$ line, we obtained the redshift and the X-ray luminosity (2--8~keV) to be $0.46^{+0.03}_{-0.04}$ and 2.6$\times10^{44}~{h_{70}}^{-2}\,\rm{erg}\,\rm{s}^{-1}$, respectively. The luminosity lies in typical values of AGNs (e.g. \cite{Silverman08}). The measured equivalent width of the Fe line ($340^{+560}_{-270}$~eV) is marginally larger than those of typical AGNs (50--200~eV; \cite{Ricci14}) although the error ranges are large. 

If the source is a CG, the 4.4-keV line would come from a Fe~{\emissiontype {XXV}} K$\alpha$ line. The spectral analysis obtained the electron temperature of $kT_{\rm e}=8.2^{+8.0}_{-2.9}$~keV. 
Using  $H_0=70~h_{70}\ \si{km\ s^{-1}.Mpc^{-1}}$, $\Omega_{\rm{M}}=0.3$, $\Omega_{\Lambda}=0.7$ and the measured redshift of $z=0.55$, the distance and the bolometric X-ray luminosity were estimated to be 2700~${h_{70}}^{-1}\,\rm{Mpc}$ and 5.3$\times10^{44}~{h_{70}}^{-2}\,\rm{erg}\,\rm{s}^{-1}$, respectively.
The electron temperature and the bolometric luminosity are within the dispersion of the $L_{\rm X}$-$kT$ plot reported by \citet{Fukazawa04}. The Fe abundance of the source ($Z=0.75^{+1.86}_{-0.47})$ is consistent with that of CGs with the Fe abundance of 0.5~solar (\cite{Fukazawa00};  \cite{Matsushita11}).
The CG origin seems to be somewhat more likely than the AGN origin. 

Since CGs and AGNs are extended and point sources, respectively, one of the ways to distinguish the two origins is to investigate the spatial extent of the source. We note that, however, even if Suzaku J1937.4$+$1718 is a CG, its extent can be indistinguishable from a point source due to the XRT's spatial resolution; assuming a typical virial radius of CGs of 1~Mpc (e.g. \cite{Walker12}), the expected apparent extent of the source is $1.'3$, which is smaller than the spatial resolution (a half-power diameter)  of the XRT of $1.'9$ (\cite{Serlemitsos07}). 

\section{Conclusion}
Analyzing the Suzaku data, we diagnosed plasma states of the SNR 3C\,400.2.
To carefully estimate the background emission, we conducted the simultaneous fitting of the spectra extracted from the whole SNR and BGD regions. In contrast to the previous studies,  which analyzed the Chandra and Suzaku data and explained the SNR spectra with the RP models, we reproduced the SNR emission by the model consisting of the low-temperature CIE plasma of the ISM origin and the high-temperature IP of the ejecta origin. The discrepancy between the previous studies and this study is considered to be due to the difference of the  background estimation methods. 

We also discovered the unidentified point-like source named Suzaku J1937.4$+$1718 near 3C\,400.2. The higher hydrogen column density than that of the SNR and the detection of the 4.4-keV line emission with the $2.8\sigma$ significance level indicate that Suzaku J1937.4$+$1718 would be an extragalactic source, either an AGN or a CG. A future study with a higher spatial resolution may distinguish its origin.

\bigskip  
\begin{ack} 
 We are grateful to Satoru Katsuda for providing his valuable comments.  This research is supported by MEXT KAKENHI No. JP20K14491, JP20KK0071, 23H00151 (KKN) and JP21K03615 (MN). KKN is also supported by the Yamada Science Foundation and the Mitsubishi Foundation.
\end{ack}


\addcontentsline{toc}{chapter}{References}
\begin{thebibliography}{99}
\bibitem[Afsariardchi et al.(2021)]{Afsariardchi21}Afsariardchi, N., Drout, M. R., Khatami, D. K., Matzner, C. D., Moon, D. -S., Ni, Y. Q. 2021, \apj, 918, 89
\bibitem[Ambrocio-Cruz et al.(2006)]{Ambrocio06} Ambrocio-Cruz, P., Rosado, M., \& de la Fuente, E. 2006, RMxAA, 42, 241
\bibitem[Anders et al.(1989)]{Anders89} Anders, E., \& Grevesse, N. 1989, Geochim. Cosmochim. Acta, 53, 197
\bibitem[Anderson(2019)]{Anderson19} Anderson, J. P. 2019, A\&A, 628, 7
\bibitem[Broersen et al.(2015)]{Broersen15}Broersen, S. \& Vink, J. 2015, \mnras, 446, 3885
\bibitem[Dubner et al.(1994)]{Dubner94}Dubner, G. M., Giacani, E. B., Goss, W. M., \& Winkler, P, F. 1994, \apj, 108, 207
\bibitem[Ergin et al.(2017)]{Ergin17}Ergin, T., Sezer, A., Sano, H., Yamazaki, R., \& Fukui, Y. 2017, \apj, 842, 22
\bibitem[Fukazawa et al.(2000)]{Fukazawa00}Fukazawa, Y., Makishima, K., Tamura, T., Nakazawa, K., Ezawa, H., Ikebe, Y., Kiuchi, K., \& Ohashi, T. 2000, \mnras, 313, 21
\bibitem[Fukazawa et al.(2004)]{Fukazawa04}Fukazawa, Y., Makishima, K., \& Ohashi, T. 2004, \pasj, 56, 965
\bibitem[Giacani et al.(1998)]{Giacani98}Giacani, E. B., Dubner, G., Cappa, C., \& Testori, J. 1998, \aaps, 133, 61
\bibitem[Ishisaki et al.(2007)]{Ishisaki07}Ishisaki, Y., et al. 2007, \pasj, 59, S113
\bibitem[Katsuda et al.(2018)]{Katsuda18}Katsuda, S., Takiwaki, T., Tominaga, N., Moriya, T. J., \& Nakamura, K. 2018, \apj, 863, 127
\bibitem[Konami et al.(2012)]{Konami12}Konami, S., Matsushita, K., Gandhi, P., \& Tamagawa, T. 2012, \pasj, 64, 117
\bibitem[Koyama et al.(2007)]{Koyama07}Koyama, K., et al. 2007, \pasj, 59, S23
\bibitem[Kushino et al.(2002)]{Kushino02}Kushino, A., Ishisaki, Y., Morita, U., Yamasaki, N. Y., Ishida, M., Ohashi, T., \& Ueda, Y. 2002, \pasj, 54, 327
\bibitem[Long et al.(1991)]{Long91}Long, K. S., Blair, W. P., White, R. L., \& Matsui, Y. 1991, \apj, 373, 567
\bibitem[Matsushita(2011)]{Matsushita11}Matsushita, K. 2011, A\& A, 527, A134
\bibitem[Mitsuda et al.(2007)]{Mitsuda07}Mitsuda, K., et al. 2007, \pasj, 59, S1
\bibitem[Nakashima et al.(2013)]{Nakasima13}Nakashima, S., Nobukawa, M., Uchida, H., Tanaka, T., Tsuru, T. G., Koyama, K., Murakami, H., \& Uchiyama, H. 2013, \apj, 773, 20
\bibitem[Ono et al.(2019)]{Ono19}Ono, A., Uchiyama, H., Yamauchi, S., Nobukawa, M., Nobukawa, K. K., \& Koyama, K. 2019, \pasj, 71, 52
\bibitem[Rho \& Petre(1998)]{Rho98}Rho, J. \& Petre, R. 1998, ApJ, 503, L167
\bibitem[Ricci et al.(2014)]{Ricci14}Ricci, C., Ueda, Y., Ichikawa, Y., Paltani, S., Boissay, R., Gandhi, P., Stalevski, M., \& Awaki, H. 2014, A\&A, 567, 142
\bibitem[Rosado(1983)]{Rosado83}Rosado, M. 1983, Rev. Mex. Astron. Astrofis., 8, 59
\bibitem[Saken et al.(1995)]{Saken95}Saken, J. M., Long, K. S., Blair, W. P., \& Winkler, P. F. 1995, \apj, 443, 231
\bibitem[Serlemitsos et al.(2007)]{Serlemitsos07}Serlemitsos, P. J., et al. 2007, \pasj, 59, S9
\bibitem[Silverman et al.(2008)]{Silverman08}Silverman, J. D., et al. 2008, \apj, 679, 118
\bibitem[Tawa et al.(2008)]{Tawa08}Tawa, N., et al., 2008, \pasj, 60, S11
\bibitem[Uchiyama et al.(2008)]{Uchiyama08}Uchiyama, Y., Maeda, Y., Ebara, M., Fujimoto, R., \& Ishisaki, Y. 2008, \pasj, 60, S35
\bibitem[Uchiyama et al.(2013)]{Uchiyama13}Uchiyama, H., Nobukawa, M., Tsuru, T. G., \& Koyama, K. 2013, \pasj, 65, 19
\bibitem[Walker et al.(2012)]{Walker12}Walker, S., Fabian, A., Sanders, J., George, M., \& Tawara, Y. 2012, AIPC, 1427, 338
\bibitem[Winkler et al.(1993)]{Winkler93}Winkler, P. F., Olinger, T. M., Ratcliff, S. J., \& Westerbeke, S. A. 1993, \apj, 405, 608
\bibitem[Yamauchi et al.(2016)]{Yamauchi16}Yamauchi, S., Nobukawa, K. K., Nobukawa, M., Uchiyama, H., \& Koyama, K. 2016, \pasj, 68, 59
\bibitem[Yamauchi et al.(2023)]{Yamauchi23}Yamauchi, S., \& Panutti, T. G. 2023, \pasj, 75, 1273
\bibitem[Yoshita et al.(2001)]{Yoshita01}Yoshita, K., Tsunemi, H., Miyata, E., \& Mori, K. 2001, \pasj, 53, 93
\end{thebibliography}
\end{document}